\RequirePackage{ifpdf}
\ifpdf 
\documentclass[pdftex]{sigma}
\else
\documentclass{sigma}
\fi

\begin{document}

\renewcommand{\PaperNumber}{022}

\FirstPageHeading

\ShortArticleName{Noether Symmetries and Critical Exponents}

\ArticleName{Noether Symmetries and Critical Exponents}

\Author{Yuri BOZHKOV}

\AuthorNameForHeading{Yu. Bozhkov}

\Address{Departamento de Matem\'atica Aplicada - DMA,  Instituto de Matem\'atica,\\
  Estatistica e
 Computa\c c\~ao Cient\'\i fica - IMECC, Universidade
Estadual de Campinas -\\
 UNICAMP, C.P. $6065$, $13083$-$970$ - Campinas - SP, Brasil}

\Email{\href{mailto:bozhkov@ime.unicamp.br}{bozhkov@ime.unicamp.br}}

\URLaddress{\href{http://www.ime.unicamp.br/~bozhkov/}{http://www.ime.unicamp.br/\~{}bozhkov/}}

\ArticleDates{Received October 03, 2005, in final form November
19, 2005; Published online November 25, 2005}

\Abstract{We show that all Lie point symmetries of various classes
of nonlinear differential equations involving critical
nonlinearities are variational/divergence symmetries.}

\Keywords{divergence symmetry; critical exponents}

\Classification{34C14; 35A15}

\section{Introduction}

The main purpose of this paper is to discuss a common property of
certain classes of quasilinear differential equations, namely:
{\it a Lie point symmetry of the considered equation is a Noether
symmetry if and only if the equation parameters assume critical
values}.

By `Noether symmetry' we mean a variational or a divergence
symmetry. Further, as it is well known, the so-called critical
exponent is found as the critical power for embedding theorems. It
is also related to some numbers dividing the existence and
nonexistence cases for the solutions of differential equations, in
particular, of semilinear differential equations with power
nonlinearities involving the Laplace operator. The above stated
property traces a connection between these two notions: the
Noether symmetries and the `criticality' of the equation.

We shall review the results obtained in \cite{yg1,yg2,yg3,yb},
where that property was established for various differential
equations. To the author's knowledge, the `enigmatic' relation
between the Noether symmetries and the critical exponents has not
previously been emphasized.

Our research was inspired by the paper of Clement, de Figueiredo
and Mitidieri \cite{cfm} in which these authors introduced the
following class of quasilinear ordinary differential equations:
\begin{gather}\label{e01}
-\big(x^{{\alpha}}|y'|^{{\beta }}y'\big)'= x^{\gamma }
f(y),\end{gather}
 where $\alpha$, $\beta $ and $\gamma $ are real numbers, $x>0$, $y=y(x)$ and
 $f$ is a non-negative function. The relations between these
 parameters will be specified below. We point out that for specific
 values of $\alpha$, $\beta$, $\gamma $ it
 contains well-known important differential equations
 briefly listed in Section~2. In particular, it contains the radial
 forms of partial differential equations involving Laplace or
 $p$-Laplace operators. Our aim is to investigate the cases, in
 terms of the equation parameters, for which the stated in the
 beginning property holds.

 We call the equation~\eqref{e01} the {\it model equation}. It is
 is considered for $x\in (0,R)$,
$0<R\leq\infty $, with the conditions
\begin{gather} \label{e02}
y'(0)=0,\qquad y(R)=0, \end{gather} and one looks for positive
solutions.

 We are interested in nonlinearities of power or exponential type.
 According to \cite{cfm} we distinguish the following two cases:

 -- The Sobolev case: $f(y)=\lambda y^{{p}} $, $\lambda={\rm const}$, and
 \begin{gather} \label{d4}
 \alpha -\beta -1 >0 ;
 \end{gather}

 -- The Pohozaev--Trudinger case: $f(y)=\lambda e^{{y}}$, $\lambda ={\rm const}$, and
 \begin{gather} \label{d5}
 \alpha =\beta +1.
 \end{gather}

If the equations involved the Laplace operator, both cases are
related with the exceptional cases of the Sobolev theorem. In this
regard we recall that the existence of solution to the equation
$\Delta u + u^p=0$ depends essentially on the number $p$
characterizing the nonlinearity. Namely, it admits positive
solutions if and only if $p\geq 2^{*}= (n+2)/(n-2)$ --- the
critical Sobolev exponent~\cite{gs}.

For the Sobolev case the critical exponent associated to (1) was
found in~\cite{cfm}. Its value is given by
 \begin{gather} \label{d7} q^{*}=\frac{(\gamma +1)(\beta +2)}{\alpha -\beta -1}. \end{gather}

Some explicit formulas for the solutions corresponding to the
above cases were found in \cite{cfm}, among other results. The
relations \eqref{d4} and \eqref{d5} were essentially used in that
work. It is natural to raise the following questions:

What is the relationship between the conditions (\ref{d4}) and
(\ref{d5}) and the nature of the equation~(\ref{e01})? In what
context do they appear? What is the role of the critical exponent?
After all, why was it possible to find the exact solutions?

We asked and answered these questions in \cite{yg1,yg2}. In those
works the Lie point symmetry group of the model equation is
calculated. Then it is verified which of the found symmetries are
Noether symmetries. In this way, the property mentioned in the
beginning
 regarding Noether symmetries and critical exponents, is
established on the base of the S.~Lie Theory. Since in the most
cases the proof of embedding theorems between function spaces is
reduced to radially symmetric functions, an ordinary differential
equation is obtained. The symmetry approach permits as well the
critical exponent  be defined using directly that ordinary
differential equation, without involving functional analysis.
Namely: it is the only exponent for which a Lie point symmetry of
(\ref{e01}) with $f(y)=\lambda y^p$ is a variational symmetry.

Further, in \cite{yg3} and \cite{yb} respectively, similar
problems are considered for systems of ordinary differential
equations and scalar partial differential equations. In this paper
we shall present the main results of \cite{yg1,yg2,yg3,yb}. Our
exposition in some parts follows closely the text of the original
articles. It corresponds to the talk we gave on June 22, 2005,
during the 6th International Conference `Symmetry in Nonlinear
Mathematical Physics',  June 20--26 2005, Kyiv, Ukraine.

\section{The model equation}

The model equation contains the following particular cases:

1. The equation (\ref{e01}) is the radial form of partial
differential equations containing
\begin{enumerate}
\itemsep=-1pt \item[--] the Laplace operator in ${ {\mathbb{R}
}}^N $ if $\alpha = \gamma =N-1$, $\beta =0$;

\item[--] the $p$-Laplace operator in ${\mathbb{R} }^N $ if
$\alpha=\gamma =N-1$, $\beta =p-2$;

\item[--] the $k$-Hessian operator in ${\mathbb{R} }^N $ if
$\alpha =N-k$, $\gamma=N-1$, $\beta = k-1$.
\end{enumerate}

2. The equation (\ref{e01}) with $\alpha = \gamma =N-1, \beta =0$
and $f(y)=\lambda
 e^y$ arises in the Liouville--Gelfand problem.

3. The generalized Lane--Emden equation of the first kind:
\begin{gather} \label{d1}
y'' + \frac{\tilde{\alpha }}{x} y' + \tilde{\beta }x^{{\nu
-1}}y^{{n}} =0,
 \end{gather}
 can be obtained from (\ref{e01}) by setting $\beta =0$, $\alpha =\tilde{\alpha }$,
 $f(y)=\tilde{\beta }y^n$, and
 $\gamma =\alpha + \nu -1$. The equation (\ref{d1}) with
 $\tilde{\alpha }=2$ is the Emden--Fowler equation, and if, in addition, $\nu
 =1$, then (\ref{d1}) is the equation proposed by Lane and studied
 in detail by Emden.

 This equation and its particular cases appear in astrophysics,
 mechanics, general relativity, various theories of gravitation,
 atomic physics and quantum mechanics.

4. The generalized Lane--Emden equation of the second kind:
\[ \label{d2} y'' + \frac{\tilde{\alpha }}{x} y' + \tilde{\beta
}x^{{\nu -1}}e^{{ny}} =0.
\]

5. The Boltzmann equation
\[ \label{d3 } -(xy')'= x\sum_{i=1}^{m} {\lambda }_i k_i e^{{-k_i
 y}},
 \]
which is used in a huge number of applications, in particular, in
the biophysics study of DNA modelled as a spiral on a cylinder .

\section{The Sobolev case for the model equation}

In this section we consider the model equation (\ref{e01}) with
$f(y)=\lambda y^{{p}} $, $\lambda >0$ and $\beta >-1$, which can
be written in the following form:
\begin{gather} \label{u19}
y''=-\frac{\alpha }{\beta +1}\frac{y'}{x}-\frac{\lambda }{\beta
+1} x^{\gamma -\alpha }|y'|^{-\beta } y^p.
\end{gather}

The symmetry group of \eqref{u19} is calculated in \cite{yg2},
where various cases, depending on some relations between the
parameters, are considered. Although that may be interesting from
the point of view of the group analysis, we shall not present the
complete results here directing the interested reader to
\cite{yg2} for further details. Rather, in order to establish the
property stated in the introduction regarding Noether symmetries
and critical exponents, we shall concentrate on the case when
\begin{gather} \label{u16}
 \alpha -\beta -1 >0
 \end{gather} and
 \begin{gather} \label{u17}
 \beta +1 <p \leq
q^{*} -1
\end{gather} where the critical exponent $q^{*}$ is given by \eqref{d7}.
In this way, for the sake of clarity and brevity, we shall avoid
some considerations which do not contribute to clarifying of that
property. Moreover, these conditions are natural, since they are
valid in the most of the interesting particular cases of the model
equation. The following theorem is a direct corollary of the
results obtained in \cite{yg2}.

\begin{theorem}  Let \eqref{u16} and \eqref{u17} hold. Then the Lie point
symmetry group of \eqref{u19} is generated by the vector field
 \begin{gather} \label{t33}
 X_1=\frac{\beta + 1 -p}{\gamma -\alpha +\beta +2}\;
 x\frac{\partial }{\partial x}+y\frac{\partial }{\partial y}.
 \end{gather} \end{theorem}

 \begin{proof}
 By \eqref{u16} and\eqref{u17} it follows that $\gamma -\alpha +\beta +2 >0$. (The latter condition
 is necessary for the existence of a positive solution of our problem
 \cite{cfm}.) It is also easy to see that \eqref{u16} and
 \eqref{u17} imply $\alpha + \gamma +\beta\gamma\neq 0$. Then the theorem
 follows from the results in \cite{yg2}. \end{proof}

 Further we observe that the equation \eqref{u19} is the Euler--Lagrange equation
 of the functional
 \[J[y]=\int_0^R\; L(x,y,y')\;dx,
 \]
 where the function of Lagrange $L=L(x,y,y')$ is given by
 \[L(x,y,y')={\frac{1}{\beta +2}}x^{\alpha }|y'|^{\beta +2}-\frac{\lambda}{p+1}\; x^{\gamma }
 y^{p+1}. \]

 Our next purpose is to check which of the symmetries leave the
 Euler functional $J[y]$ invariant, that is, which of the already
 found in~\cite{yg2} Lie point symmetries are actually variational.
 Again we consider only the case in the Theorem 1 in which all
 symmetries form a one-parameter group generated by $X_1$.

 \begin{theorem}[\cite{yg2}] A Lie point symmetry of the equation \eqref{u19} is a
 variational symmetry if and only if $p+1$ is equal to the critical
 exponent.
 \end{theorem}

 \begin{proof} We denote
 \[ k= \frac{\beta + 1 -p}{\gamma -\alpha +\beta +2}.
 \]
 The Lie point transformation corresponding to \eqref{t33} is given by
  \begin{gather*}
  x^{*} = {\mu }^k x, \\
  y^{*} = \mu y .
  \end{gather*}
  Clearly this is a scaling transformation. Then
\begin{gather*}
 J[y^{*}] =  \frac{1}{\beta +2}\int_0^{R^{*}} (x^{*})^{\alpha }
  \left | \frac{dy^{*}}{dx^{*}}(x^{*})\right |^{\beta +2}dx^{*} - \frac{\lambda
}{p+1}
  \int_0^{R^{*}} (x^{*})^{\gamma } (y^{*}(x^{*}))^{p+1}dx^{*}\\
\phantom{J[y^{*}]}=\frac{{\mu }^{k(\alpha -\beta -1)+\beta
+2}}{\beta +2}
  \int_0^R x^{\alpha }|y'|^{\beta +2}dx-\frac{\lambda{\mu }^{k(\gamma
  +1)+p+1}}{p+1}\int_0^R x^{\gamma }y^{p+1} dx.
\end{gather*}
  Hence $J[y^{*}] = J[y]$ if and only if both exponents of $\mu $ vanish:
\begin{gather*}
 k(\alpha -\beta -1)+\beta +2=0, \\
 k(\gamma +1)+p+1 =0.
 \end{gather*}
The latter two equalities hold if and only if $p=q^{*}-1$.
\end{proof}

 Henceforth, to the end of this section, we shall suppose $p=q^{*}-1$. Then by
 the Noether theorem we obtain that
 \begin{gather} \label{u20}
 {\frac{\beta +1}{\alpha -\beta -1}} x^{{\alpha +1}}
 |y'|^{{\beta +2}}-x^{\alpha }y|y'|^{{\beta +1}}+\frac{\lambda }{\gamma +1} x^{{\gamma +1}}
 y^{{q^{*}}}=0,
 \end{gather}
 whenever $y$ is a solution of the equation (\ref{u19}).

 Taking a look at
 \cite{cfm}, one can observe that the above first integral, up to some
 constant multiplier, is the
 $\psi $ used in \cite[p.~154]{cfm}. Further we
 express $\lambda x^{\gamma -\alpha} y^{q^{*}-1}$ from (\ref{u20}) as a function of
 $x$ and $y'$ and then substitute it into the equation (\ref{u19}). We thus
 obtain that $y'$ must
 satisfy the following equation
 \[
 y''= \frac{(\gamma - \alpha +1)}{\beta +1 }\frac{1}{x}y' +
 \frac{\gamma +1}{\alpha -\beta -1 }\frac{1}{y}{y'}^2,
 \]
 whose solution, represented below, can be easily found
 by setting $y'=vy$.
 In this way we have come to
 the following theorem of \cite{cfm}:

\begin{theorem}[\cite{cfm}] The problem
 \begin{gather} \label{u224}
 \begin{array}{lll}
-\big(x^{\alpha}|y'|^{\beta }y'\big)'= \lambda\;x^{\gamma }
y^{q^{*}-1} & \mbox{\rm \ in\ } & (0,\infty ) ,
\vspace{1mm}\\
 y(0)=y_0>0,\quad y'(0)=0, &  &  \vspace{1mm}\\
 y>0 & \mbox{\rm \ in\ } & [0,\infty ) ,
 \end{array}
\end{gather}
 has the unique solution
 \[ \label{u21}
 y_{*}(x) = \left ({y_0}^{-{\sigma }} +
 k_0 x^{{s}} \right )^{{-1/\sigma}}, \]
 where
 \[\sigma ={\frac{\gamma -\alpha +\beta +2}{\alpha -\beta -1}}, \qquad
  s={\frac{\gamma -\alpha +\beta +2}{\beta +1} }
\qquad  \mbox{\rm and}\qquad
 k_0={\frac{\beta +1}{\alpha -\beta -1}}
 \left ({\frac{\lambda {y_0}^{{\sigma }}}{\gamma +1}} \right )^{1/(\beta +1)}.
 \]
 \end{theorem}

The reason for the success in solving \eqref{u224} is fact
 that the exponent in the right-hand side of the equation is
 critical, and hence any Lie point symmetry is variational, which
 reduces the order of integration procedure by two. In particular,
 this explains why the Lane--Emden equation
 \[ \label{a2} y'' +{\frac{2}{x}}\,y' + y^{{5} }=0 \]
 can be explicitly solved. Indeed, if the spatial dimension $n=3$,
 then the corresponding equation parameters are $2=n-1$ and
 $5=(n+2)/(n-2)$ --- the Sobolev exponent.

\section[The Pohozaev-Trudinger case for the model equation]{The Pohozaev--Trudinger case for the model equation}

 \rm Denote by $B$ the sphere in ${\mathbb{R} }^N$ with centre at the origin and
 radius $R$. Then the Liouville--Gelfand problem consists of finding positive
 solutions of the following equation
 \begin{gather}\label{s11}
 -\Delta u= \lambda\, e^{{u}}  \quad \mbox{\rm in} \quad  B  \end{gather}
 with the Dirichlet boundary condition:
 \begin{gather} \label{s22} u=0  \quad \mbox{\rm on}  \quad \partial B. \end{gather}
The equation above has been proposed and studied by Liouville. He
found an exact solution in dimension $N=1$, and also (for $N=2$) a
solution in terms of an arbitrary harmonic function.
Bratu~\cite{b} found two explicit solutions of the problem if
$0<\lambda < 2/R^2 $ and $N=2$. Gelfand~\cite{g} considered, among
other things, the problem of thermal self-ignition of a chemically
active mixture of gases in plane, cylindrical and spherical
vessels. For $N=3$ he investigated the values of the parameter
$\lambda $ for which the problem has a solution and studied the
multiplicity of such solutions.

It can be proved that if $\lambda \leq 0 $ or if $\lambda $ is
greater than a certain positive constant ${\lambda }^{*}$ there is
no solution of the boundary value problem \eqref{s11},
\eqref{s22}. Then by the celebrated results of Gidas, Ni and
Nirenberg \cite{gnn} it follows that if a solution exists, then it
must be radially symmetric. Thus the Liouville--Gelfand problem is
reduced to the study of the following problem for an ordinary
differential equation:
 \[
\begin{array}{lll}
\displaystyle -y''-\frac{N-1}{x}y'= \lambda\, e^{{y}} & \mbox{\rm
\ in\ } & (0,R) ,
\vspace{1mm}\\
 y'(0)=y(R)=0, &  &  \vspace{1mm}\\
 y>0 & \mbox{\rm \ in\ } & [0,R) .
 \end{array}
 \]

 In \cite{cfm} Clement, de Figueiredo and Mitidieri  proposed the
following generalization : \begin{gather} \label{f3}
\begin{array}{lll}
-\big(x^{\alpha}|y'|^{\beta }y'\big)'= \lambda\;x^{\gamma }
e^{{y}} & \mbox{\rm \ in\ } & (0,R) ,
\vspace{1mm}\\
 y'(0)=y(R)=0, &  &  \vspace{1mm}\\
 y>0 & \mbox{\rm \ in\ } & [0,R) ,
 \end{array}
 \end{gather} assuming that \[
  \alpha -\beta -1=0,\qquad \beta > -1, \qquad \gamma >-1,
  \]
that is, the situation corresponds to the Pohozaev--Trudinger
case.

Clement, de Figueiredo and Mitidieri  found in  \cite{cfm} the
above mentioned constant ${\lambda }^{*}$, proved that there is
only one solution of \eqref{f3} if $\lambda ={\lambda }^{*}$, and
that there exist exactly two solutions if $0<\lambda < {\lambda
}^{*}$. Moreover, using the method of first integrals, they found
the explicit formulas for these solutions. Again the reason for
the success of the integration is the fact that the condition
$\alpha =\beta +1 $ used in \cite{cfm} holds if and only if all
Lie point symmetries are variational symmetries, and hence the
order of the integration procedure reduces by two. This is proved
in \cite{yg1} where it is shown that the exact solutions can be
accounted for by symmetry techniques. To see this, one needs first
to find the Lie point symmetries of the model equation with
$f(y)=\lambda e^u$, $\lambda >0$. We present the result in the
following reduced form

\begin{theorem}[\cite{yg1}] Suppose that $\beta >-1$, $\alpha -\beta -1\geq 0$ and
$\gamma -\alpha +\beta +2>0$. If $\beta\neq 0$ or if $\beta =0$,
$\alpha\neq 1$, the Lie point symmetry group of the equation
\[-\big(x^{\alpha}|y'|^{\beta }y'\big)'= \lambda\;x^{\gamma } e^{{y}} \]
is generated by
\[X_2=-\frac{1}{m}\,
 x\frac{\partial }{\partial x}+\frac{\partial }{\partial y}, \]
 where $m=\gamma -\alpha +\beta +2$. If $\beta =0$ and $\alpha =1$ the symmetry group is
 the
 two-parameter Lie group determined by $X_2$ and
 \[ {\frac{1}{\gamma +1}}\left(
 {\frac{2}{\gamma +1}}\, x -x\ln x \right)\frac{\partial }{\partial x}
 +\ln x\frac{\partial }{\partial y}.\]
 \end{theorem}

 In the quasilinear case, it is easy to see by Theorem 4
 that the following result holds.

  \begin{theorem}[\cite{yg1}] Let $\beta\neq 0$. Then the Lie point symmetry group of
  \[ -\big(x^{\alpha}|y'|^{\beta }y'\big)'= \lambda\, x^{\gamma } e^{{y}}
 \]
is a variational symmetry group if and only if $ \alpha =\beta
+1.$
\end{theorem}

 Further, observe that by Theorem 4, the Lie point symmetry group for $\beta =0$ and $\alpha =1$
 is two-dimensional. Hence, one can deduce that in this case there exists a
 one-dimensional subgroup consisting of variational symmetries.

 Then using $\alpha =\beta +1$ and $|y'|=-y'$,  we
 obtain the following first integral
 \begin{gather}\label{o11}
 \psi (x,y,y'):=\alpha x^{\alpha +1} |y'|^{\alpha +1}-(\alpha +1)(\gamma +1)x^{\alpha }|y'|^{\alpha }
 + \lambda (\alpha +1) x^{\gamma +1} e^{{y}}=0.
 \end{gather}
 Again, taking a look at \cite{cfm}, one can observe that the function
 $\psi $ above is the same that appears in \cite{cfm}. Further we
 express $y$ from \eqref{o11} as a function of $x$ and $y'$ and then
 substitute it into the equation in \eqref{f3}. We obtain that $y'$ must
 satisfy the following Bernoulli equation
 \[y''= \frac{(\gamma - \alpha +1)}{\alpha }\frac{1}{x}y' + \frac{1}{\alpha +1 }{y'}^2
 \]
 which can be easily solved. Thus
 \[y(x)= - (\alpha +1)\ln \left | c_1 - \frac{\alpha }{(\alpha +1)(\gamma +1)}
 x^{(\gamma +1)/\alpha } \right | + c_2.\]
 By the boundary condition $y(R)=0$ (see (2)) we obtain that \[c_1=-
 \frac{(\gamma +1)^{\alpha }}{{\mu }_i (\lambda {\mu }_i)^{1/\alpha }}\qquad \mbox{\rm and}\qquad
 c_2 = (\alpha +1)\ln \frac{(\gamma +1)^{\alpha }}{(\lambda {\mu }_i)^{1/\alpha }},
 \] where ${\mu }_i$, $i=1,2$, is a root of
 \[ H(\mu ):= m {\mu }^{(\alpha +1)/\alpha }-\mu +1 =0,\qquad
 m=\frac{\alpha {\lambda }^{1/\alpha } R^{(\gamma+1)/\alpha }}
 {(\alpha +1)(\gamma +1)^{(\alpha +1)/\alpha  }}. \]

 In this way we have come to another result of \cite{cfm}:

\begin{theorem}[\cite{cfm}] The solution(s) of $\eqref{f3}$ can be represented in the following way:

$1)$ if $0<\lambda <{\lambda }^{*}:={\frac{(\gamma +1)^{\alpha
+1}}{(\alpha +1)R^{\gamma +1}}}$ then there are two solutions \[
y_{1,2}(x) =-(\alpha +1) \ln \left ( \frac{1}{{\mu }_{1,2}} +
\frac{\alpha (\lambda {\mu }_{1,2})^{1/\alpha } }{(\alpha
+1)(\gamma +1)^{(\alpha +1)/\alpha }} \; x^{(\gamma +1)/\alpha
}\right ); \]

$2)$ if $ \lambda ={\lambda }^{*}$ then there exists the unique
solution
\[
y_0(x)=-(\alpha +1) \ln \left ( \frac{1}{{\mu }_{0}} +
\frac{\alpha {{\mu }_0}^{1/\alpha }}{(\alpha +1)^{(\alpha
+1)/\alpha }R^{(\gamma +1)/\alpha }} \; x^{(\gamma +1)/\alpha
}\right ), \] where ${\mu }_0$ is the unique solution of $H(\mu
)=0$;

$3)$ if $\lambda \leq 0$ or $\lambda > {\lambda }^{*}$ there is no
solution.\end{theorem}

We observe that we have obtained the solutions found in \cite{cfm}
just by applying
 symmetry and variational methods, and without use of another first integral
\[
\varphi := x^{(\sigma -\gamma -1)/\alpha } \frac{d}{dx}\left (
e^{-y(x)/(\alpha +1)}\right ).
\]
 See \cite{cfm}. In fact, this $\varphi
$ corresponds to a {\it dynamical} symmetry.

\section{The variational symmetries of the Lane--Emden systems}

In Sections 3 and 4, following \cite{yg1,yg2}, we have described
the properties of the large class of quasilinear ordinary
differential equations~\eqref{e01} with power or exponential
nonlinearities from the point of view of the Lie symmetry theory.
We have shown that the Lie point symmetries of equations which
contain critical exponents are actually variational symmetries. In
this section we shall present the results of \cite{yg3} which show
that this property is also valid for systems.

The system of two semilinear partial differential equations
\begin{gather}
-\Delta u=v^q, \nonumber\\
-\Delta v=u^{p},
 \label{400}
\end{gather}
where the independent variable $x\in  {\mathbb{R}}^{n}$, $n\geq 3$
--- integer, $p$ and $q$ --- positive real numbers, is called the {\it
Lane--Emden system}. It can be considered as a natural extension
of the celebrated Lane--Emden equation
\begin{gather}\label{j11}
\Delta \theta +\theta ^{p}=0
\end{gather}
in ${\mathbb{R}}^n$.

We are interested in \it radial ground state solutions \rm of
(\ref{400}), whose study is reduced to the study of the following
system of two ordinary differential equations:
\begin{gather}
\ddot{u}  +  \frac{n-1}{t}\dot{u}+v^{q}  =  0, \nonumber\\
\ddot{v}  +  \frac{n-1}{t}\dot{v}+u^{p}  =  0,
 \label{nccc}
\end{gather}
where $t=|x|$, $\dot{u}=\frac{du}{dt}$, etc., and $\dot{u}( 0)
=\dot{v}( 0) =\lim\limits_{t\rightarrow \infty }u( t)
=\lim\limits_{t\rightarrow \infty }v( t) =0$.

It is well known that the Lane--Emden equation (\ref{j11}) admits
positive solutions if and only if $p$ is greater than or equal to
the Sobolev exponent $( n+2) /( n-2) $ (see, for instance,
\cite{gs}). In~\cite{sz2} J.~Serrin and H.~Zou prove an analogous
result for the system (\ref{400}). Namely: if $( p,q) $ is on or
above the curve in the first quadrant in the $(p,q)$ plane given
by
\begin{gather}
\frac{n}{p+1}+\frac{n}{q+1}=n-2,  \label{nc}
\end{gather}
then there exist infinitely many (componentwise) positive radial
solutions $(u,v)$ of \eqref{400} tending to $(0,0)$ when
$|x|\rightarrow\infty $. That is, the solutions are ground states.
This theorem combined with the earlier nonexistence results of
E.~Mitidieri \cite{er,em1} and J.~Serrin, H.~Zou \cite{sz3,sz1}
gives a complete picture of existence and nonexistence of positive
radial solutions of (\ref{400}) in which the dividing curve
(\ref{nc}) plays an important role (see \cite[Corollary~1.2,
p.~370]{sz2}). For this reason (\ref{nc}) is called the {critical
hyperbola}.

In regard to existence of \it ground states \rm for the
Lane--Emden system \eqref{400}, the results of
\mbox{P.-L.~Lions~\cite{l}} and J.~Hulshof, R.~van der
Vorst~\cite{hv} imply that the problem with $p$ and $q$ satisfying
(\ref{nc}) has a unique, up to scalings and translations, ground
state, which is positive, radially symmetric and decreasing in the
radial component $t=| x| $. We shall come back to this point
later.

The basic assumptions on the parameters are:
\begin{gather}
n\geq 3,\qquad pq>1,\qquad p\neq q,\qquad p\neq 1,\qquad q\neq 1.
\label{u4}
\end{gather}
Then the symmetry group of the Lane--Emden system \eqref{nccc} is
calculated in \cite{yg3}, and we state it in the form of the
following
\begin{theorem}[\cite{yg3}] Suppose that the conditions \eqref{u4} hold.
Then the Lie point symmetry group of the Lane--Emden system
\eqref{nccc} is the one-parameter group generated by the vector
field
\[
X_3=t\frac{\partial }{\partial t}+\frac{2( 1+q) }{1-pq}u%
\frac{\partial }{\partial u}+\frac{2( 1+p) }{1-pq}v\frac{%
\partial }{\partial v}.
\]
\end{theorem}

Further, following the same argument as in the proof of Theorem~2,
we obtain the next

\begin{theorem}[\cite{yg3}] Let the parameters $n$, $p$ and $q$ satisfy
\eqref{u4}. Then any Lie point symmetry of the Lane--Emden system
\eqref{nccc} is a variational symmetry if and only if
\[
\frac{n}{2}=\frac{( p+1) ( q+1) }{pq-1},\] that is, the point $(
p,q) $ is on the critical hyperbola $(\ref{nc}) $. \end{theorem}

 The statement of Theorem~8 is a conjecture of Enzo Mitidieri~\cite{em2}.

We observe that if $(p,q)$ belongs to the critical hyperbola, the
Noether Theorem immediately gives the first integral used in
\cite{hv} and called invariant paraboloid.

We conclude this section noting that the exact solution of the
problem \eqref{nccc} is not available. As mentioned above, it is
known to exist, and only the asymptotic behavior of such ground
state is known. Challenged by Enzo Mitidieri and provoked by
J.~Hulshof, R. van der Vorst's paper we proposed in \cite{yg3} the
following

\smallskip

\noindent {\bf Problem.} \rm Find the solution of the Lane--Emden
system explicitly.

\section{The Lie point symmetries of semilinear polyharmonic\\ equations}

In the next section we shall consider the semilinear polyharmonic
equation
 \begin{gather}\label{e55} (-1)^{ k}{\Delta }^{ k} u = f(u), \end{gather}
 where $\Delta $ is the Laplace operator in
 ${\mathbb{R} }^n$, $n\geq 2$ and $k\geq 1$. The Lie point symmetries
 of \eqref{e55} were studied by Svirshchevskii,
 \cite{sv2,sv3} who obtained the complete group classification. We shall
 cite partially his result in a form suitable for our purpose to
 investigate the cases of power and exponential nonlinearities.

\begin{theorem}[\cite{sv2,sv3}] \label{theorem12} The widest Lie point symmetry group
admitted by \eqref{e55} with
 general $f(u)$ is determined by translations and rotations.

 For some special choices of the right-hand side $f(u)$ it can be
 expanded by additional operators as follows:

 If $f(u)=u^{ p} $, $p\neq 0$, $p\neq
 1$, we have the generator of dilations
 \[
 Z=x_i \,\frac{\partial }{\partial x_i} + \frac{ 2k}{
 1-p}\, u\, \frac{\partial }{\partial u} ,
 \]
 and, for $p=(n+2k)/(n-2k)$, there are $n$ additional generators
 \[
 Y_i =\big(2x_i x_j -|x|^2{\delta }_{ij} \big)\,\frac{\partial }{\partial
 x_j}+(2k-n)\, x_i\, u\frac{\partial }{\partial u}.
 \]

 Further, if $f(u)=e^{ u}$ then the operator
 \[
 W=x_i\; \frac{\partial }{\partial x_i} - 2k \; \frac{\partial }{\partial u}
 \]
 generates a sub-group of the Lie point symmetry group of $(\ref{e55})$. If
 $n=2k$, there are $n$ additional generators
 \[
 V_i =\big(2x_i x_j -|x|^2{\delta }_{ij} \big)\,\frac{\partial }{\partial
 x_j}-4k\, x_i\, \frac{\partial }{\partial u}.
 \]
\end{theorem}
 \rm Above ${\delta }_{ij}$ is the Kronecker symbol,
 $|x|=\Big(\sum\limits_{s=1}^{n}x_s^2\Big)^{1/2}$ and summation over a repeated index is
 assumed.

\section{The Noether symmetries of semilinear polyharmonic\\ equations with
critical nonlinearities}

 Up to this point, we have established the property relating the
 critical exponents and the Noether symmetries for some ordinary
 differential equations and systems. It is natural to conjecture that
 the Lie point symmetries of more
 general differential equations involving critical exponents are Noether
 symmetries. In \cite{yb} we show that this is valid for a class of
 partially differential equations. Namely, for the semilinear
 polyharmonic equation \eqref{e55}.

 Again, we shall distinguish two cases: we have
 \begin{gather}\label{s44}
 f(u)=u^{ p},\qquad n>2k\geq 2,\qquad  p\neq 0, \qquad p\neq1,
 \end{gather}
 in the Sobolev case, while in the Pohozaev-Trudinger case
 \[
 f(u)=e^{ u},\qquad n=2k, \qquad k\geq 2.
 \]

 Having at our disposal the already cited group classification (see the preceding section),
 we investigate which of the Lie
 point symmetries in these cases are
 variational or divergence symmetries.

 The first results in \cite{yb} can be formulated as follows:

 \begin{theorem}[\cite{yb}] Suppose that the conditions $\eqref{s44}$ hold. Then
 the Lie point symmetry $Z$ (see Theorem $9$) of the polyharmonic equation
 \[
 (-1)^{ k}{\Delta }^{ k} u = u^{ p} \]
 is a variational symmetry if and only if
 \[
 p=\frac{n+2k}{n-2k},
 \]
 that is, $p$ is equal to the well known critical
 exponent.\end{theorem}

\begin{theorem}[\cite{yb}] Let $n\geq 3$ and $k\geq 1$. Then
 the Lie point symmetry $W$ (see Theorem $9$) of the polyharmonic equation
 \[
 (-1)^{ k }{\Delta }^{ k} u = e^{ u}.
 \]
 is a variational symmetry if and only if
 \[
 n=2k.
 \]
\end{theorem}

\rm The importance of the relation $n=2k$ for the polyharmonic
equation (\ref{e01}) was pointed out to us by Enzo Mitidieri in
January 1995~\cite{em}.

The proofs of Theorems 10 and 11 are analogous to the proof of
Theorem 2. In fact, this is a dimensional analysis argument since
the latter is equivalent to the invariance under scaling
transformations (see, for example, \cite{bk,ol}).

Since the translations and the rotations are variational
symmetries of the considered polyharmonic equation, it follows
from Theorems 10 and 11 that in the critical cases the
corresponding Lie point symmetry groups contain a subgroup
consisting of variational symmetries. However, in the critical
cases one can get more.

 The main results in \cite{yb} are the following

\begin{theorem}[\cite{yb}] Suppose that the conditions $\eqref{s44}$ hold. Then
 any Lie point symmetry of the polyharmonic equation
 \[
 (-1)^{ k} {\Delta }^{ k} u = u^{ (n+2k)/(n-2k)}
 \]
 is a divergence symmetry. \end{theorem}

 \begin{theorem}[\cite{yb}]
  Let $n\geq 4$, $k\geq 2$ and $n=2k $. Then
 any Lie point symmetry of the polyharmonic equation
 \[
 (-1)^{ k} {\Delta }^{ k} u = e^{ u}
 \]
 is a divergence symmetry.
 \end{theorem}

 The proofs of Theorems 12 and 13 are reduced to the proof of the facts
 that $Y_i$ and $V_i$ (see Theorem 9) are divergence symmetries, since the other infinitesimal
 generators of the corresponding Lie point symmetry groups determine
 variational symmetries by Theorems 10 and 11 as already observed. For
 this purpose we find explicitly
 the vector-valued `potential' function $B$ determining $Y_i$ and $V_i$
 as divergence symmetries~\cite{yb}.

 \rm The importance of the variational and the divergence symmetries is due to the
 fact that they determine conservation laws via the Noether
 Theorem \cite{bk,ol}. Thus the next step in this research is to
 establish the conservation laws corresponding to the already
 studied variational and divergence symmetries. This is possible since we already have at
 our disposal the explicit formula for $B$. This is also done in
 \cite{yb}.

 The Theorems 10--13 confirm the validity of the general property
 relating Noether
 symmetries and the critical parameters, stated in the beginning of the introduction. Another
 examples which illustrate this property are given in~\cite{sv1}, in which
 Svirshchevskii proved that the
 symmetries of the $p$-Laplace equation
 \[
 -{\rm div}\big(|\nabla u|^{p-2}\nabla u\big)=|u|^{q-1}u
 \]
 and the equation
\[
 -\Delta \big(|\Delta u|^{\sigma }\Delta u\big)=|u|^{q-1}u
 \]
 are variational if and only if the parameters assume critical
 values.

\subsection*{Acknowledgements}

We wish to thank the Organizers of the 6th International
Conference `Symmetry in Nonlinear Mathematical Physics', June
20--26
 2005, Kyiv, Ukraine, for having given us the opportunity to
talk on this subject as well as for their warm hospitality. We are
grateful to the referees for their useful suggestions. We would
also like to thank FAEPEX-UNICAMP for partial financial support.

\LastPageEnding

\end{document}